\documentclass[prl, reprint, superscriptaddress,showpacs,floatfix]{revtex4-2}
\usepackage{amsmath,amssymb, dsfont}
\usepackage{graphicx}
\usepackage{bm}
\usepackage[usenames,dvipsnames,svgnames,table]{xcolor}
\usepackage[colorlinks=True,linkcolor=blue,citecolor=blue,urlcolor=blue]{hyperref}
\usepackage{braket}
\usepackage{bbm}
\usepackage{soul}

\let\oldciteauthor=\citeauthor
\def\citeauthor#1{\hypersetup{citecolor=blue}\oldciteauthor{#1}}

\let\oldciten=\onlinecite
\def\onlinecite#1{\hypersetup{citecolor=blue}\oldciten{#1}}

\let\oldcite=\cite
\def\cite#1{\hypersetup{citecolor=blue}\oldcite{#1}}

\newcommand*\diff{\mathop{}\!\mathrm{d}}
\newcommand{\pos}{\textbf{r}}

\begin{document}
\title{Local Topological Markers in Odd Spatial Dimensions and Their Application to Amorphous Topological Matter}

\author{Julia D.\ Hannukainen}\thanks{J. D. H. and M. F. M contributed equally to this work.}
\
\author{Miguel F.\ Martínez}\thanks{J. D. H. and M. F. M contributed equally to this work.}
\
\author{Jens H.\ Bardarson}
\
\author{Thomas Klein Kvorning}
\affiliation{Department of Physics, KTH Royal Institute of Technology, 106 91, Stockholm, Sweden}
\

\begin{abstract}
Local topological markers, topological invariants evaluated by local expectation values, are valuable for characterizing topological phases in materials lacking translation invariance. 
The Chern marker---the Chern number expressed in terms of the Fourier transformed Chern character---is an easily applicable local marker in even dimensions, but there are no analogous expressions for odd dimensions. 
We provide general analytic expressions for local markers for free-fermion topological states in odd dimensions protected by local symmetries:
a \textit{Chiral marker}, a local $\mathbb Z$ marker which in case of translation invariance is equivalent to the chiral winding number, and a \textit{Chern-Simons marker}, a local $\mathbb Z_2$ marker characterizing all nonchiral phases in odd dimensions.
We achieve this by introducing a one-parameter family $P_{\vartheta}$ of single-particle density matrices interpolating between a trivial state and the state of interest. 
By interpreting the parameter $\vartheta$ as an additional dimension, we calculate the Chern marker for the family $P_{\vartheta}$.
We demonstrate the practical use of these markers by characterizing the topological phases of two amorphous Hamiltonians in three dimensions: a topological superconductor ($\mathbb Z$ classification) and a topological insulator ($\mathbb Z_2$ classification).
\end{abstract}

\maketitle
\textit{Introduction.}---Topological invariants are important for characterizing topological phases of matter~\cite{moessner_moore_2021}.
The topological invariants of free-fermion phases of crystalline solids are known: they are momentum space expressions that rely on translation invariance, a prominent example being the Chern number, the $\mathbb{Z}$-invariant of phases in even spatial dimensions~\cite{thouless82,kohmoto85,read00}.
These invariants are no longer available for characterizing the topological phases of structures far from a translationally invariant limit, such as amorphous topological matter~\cite{ Agarwala2017, Mansha2017, Mitchell2018, Ojanen2018, YanBin2019, Ojanen2020, Ojanen2020, Agarwala2020, Marsal2020, Spring2021, Grushin2022}.
Amorphous structures are easier to grow experimentally than perfect crystals, so understanding their topological properties is of interest both for technological and theoretical purposes~\cite{Zallen1998, Grushin2020}.
Therefore it is a relevant task to characterize their topological phases, which emphasizes the importance of real space formulations for topological invariants.
Such real-space characteristics are collectively denoted as topological markers~\cite{Kitaev20062, Prodan2010, Prodan2011, Loring2010, bianco11, LORING2015, Huang2018,Loring2019, Hugues2019,Irsigler2019,Jezequel2022}, and have been used for identifying topological phases in quasicrystalline~\cite{Huang2018,Huang20182}, disordered~\cite{Mondragon-Shem2014,Hughes2021,Skipetrov2021} and amorphous~\cite{Agarwala2017,Marsal2020,Agarwala2020,Focassio2021,Wang2022} settings.
Topological markers have mainly been defined for two-dimensional systems, including the local Chern marker~\cite{Kitaev20062,bianco11} and the nonlocal Bott index~\cite{Loring2010}, which are real space realizations of the Chern number, and the spin Bott index~\cite{Huang2018}, equivalent to the $\mathbb{Z}_2$ invariant~\cite{FuLiangKane2006,Liang2007} of the quantum-spin-Hall state~\cite{KaneMele20051} in two dimensions.

Topological markers in three dimensions are scarce. 
Examples include a marker for the odd-dimensional topological insulators with a $\mathbb{Z}$ invariant in the complex topological classes~\cite{Mondragon-Shem2014}, and an extension of the spin Bott index to three dimensions~\cite{Focassio2021}, 
but there exists no general local marker for odd-dimensional topological insulators and superconductors. 
In fact, characterizing the three-dimensional free-fermion topological phases for systems lacking translational invariance remains a hard task with few numerical options.
One can either check for gapless boundary modes, by exploring if the existence of the ground-state energy gap depends on whether the boundary conditions are open or closed~\citep{Agarwala2017}, or calculate the magnetoelectric response~\cite{Malashevich2010,Sinisa2011,Olsen2017,Rauch2018, Barnett2021}.
Another option is to produce a topological phase diagram by measuring the Witten effect~\cite{Witten1979, Rosenberg2010} through the amount of charge bound by a magnetic monopole as a function of the parameters of the model~\cite{Mukati2020}.
Apart from being practically challenging, neither of these diagnostics directly predict the phase of a given state since they require comparing states in different physical settings.

In this work we present a solution to this problem by providing an analytic expression for a local topological marker in odd spatial dimensions.
Specifically, we extend the formulation of the Chern marker to odd dimensions by introducing a one-parameter family of projectors traversing between the trivial and topological state, and where the parameter acts as an additional dimension.
The general expression for the marker can be used as a real space expression for both the chiral winding number, a $\mathbb{Z}$ invariant of topological phases in odd dimensions, and the $\mathbb{Z}_2$ Chern-Simons invariant.
We use our real-space marker to characterize the topological phases of two different three-dimensional amorphous Hamiltonians: a topological insulator with time-reversal invariance, characterized by the $\mathbb{Z}_2$ invariant $\theta$-term (class AII), and a topological superconductor with both chiral symmetry and time-reversal invariance, characterized by a chiral winding number $\nu\in\mathbb{Z}$ (class DIII).

\textit{The local Chern marker as a bundle invariant.}---Two localized Slater determinant states belong to the same symmetry-protected topological phase if they can be transformed into one another through smooth transformations while preserving localization~\footnote{Localized means that all correlations decay exponentially with distance.} and the given symmetry, or by adding or removing atomic-limit bands~\cite{kitaev09}.
The single-particle density matrix $\rho$ obtained from a Slater determinant---or in the presence of a local symmetry, each symmetry-resolved block of $\rho$---is a projector onto the space of occupied states.
Given translation invariance, $\rho$ is block-diagonal in momentum space and is a smooth function of momentum.
One can therefore associate a complex vector space, the image of $\rho(\mathbf{k})$, to each momentum $\mathbf{k}$ in a smooth way, thereby obtaining an equivalence relation to a complex vector bundle.   
The topological classification of Slater determinant states therefore translates into a classification of vector bundles~\cite{kitaev09}.
The blocks of $\rho$ fulfill one of ten inequivalent constraints (AZ classes)~\cite{cartan26,zirnbauer96,altland97}, leaving ten different cases to classify.

The Chern numbers (one for each even dimension) are quantized bundle invariants that in even dimensions characterize all topological phases in three out of the five nontrivial AZ classes~\cite{ryu10}.
In real space and in terms of $\rho$, the Chern numbers~\cite{Chern1946,nakahara18}, now referred to as local Chern markers~\cite{Supplemental-Material-arxiv}, take the form
\begin{equation}
	\mathcal C(\mathbf{r})=\sum_{\alpha}\frac{\varepsilon^{i_{1},\dotsc,i_{D}} [\rho  	X_{i_{1}}\rho X_{i_{2}}\dotsb X_{i_{D}}\rho]_{(\mathbf r,\alpha),(\mathbf r,\alpha)}}{(D/2)!/(2\pi i)^{D/2}},
\label{eq:local-chern-marker}
\end{equation}
where $\varepsilon$ is the Levi-Civita symbol, $D$ is the even dimension, and the repeated indices $i$ are summed over.
$X_{i}$ is the $i$th position operator: $(X_i)_{(\mathbf r,\alpha),(\mathbf r^\prime,\beta)}=x_i \delta_{\alpha,\beta}\delta_{\mathbf r,\mathbf r^\prime}$, where $x_i$ is the i'th component of the position $\mathbf r$, and $\alpha,\beta$ denote local quantum numbers.

Without translation invariance, the relation to vector bundles is lost (since $\rho$ is no longer block-diagonal in momentum space) and the Chern numbers are no longer defined. 
The average of $\mathcal C(\mathbf r)$ over large regions is, however, still a topological characteristic.
The reason for this is that the states under consideration always have a translation-invariant long-wavelength limit, and one can therefore define Chern numbers characterizing the topological phase by the coarse-grained single-particle density matrix defined by the asymptotic behavior of $\rho$ in this limit. 
When averaging $\mathcal C(\mathbf{r})$ over larger and larger regions, it will therefore approach a well-defined translation-invariant limit. 
These averaged Chern markers only depend on the long-wave-length properties of $\rho$, so they are evaluated by
replacing $\rho$ in Eq.~\eqref{eq:local-chern-marker} by its translation-invariant coarse-grained version---the averaged markers therefore \emph{are} the coarse-grained Chern numbers characterizing the phases.
This amounts to calculating the Chern marker locally for each point in the lattice and averaging over a large enough region, such that the coarse-grained $\rho$ is effectively translation invariant.

The Chern marker is only defined in even dimensions, so constructing a local marker in odd dimensions is nontrivial.
There does exist a bundle invariant in odd dimensions called the Chern-Simons invariant~\cite{nakahara18}, but it is the modulo 1 part of a basis dependent expression and cannot be expressed as a function of the single particle density matrix alone.
This means that the Chern-Simons invariant cannot be expressed as a sum of local expectation values, so it is not a local marker.

\textit{The local Chiral marker.}---In odd dimensions three out of five nontrivial AZ classes are characterized by the chiral winding number $\nu\in\mathbb{Z}$~\cite{ryu10}.
For these classes the single-particle density matrix $\rho$ obeys a chiral constraint; there exists a real Hermitian matrix $S$ squaring to identity such that $\{\rho,S\}=S$. 
We define a local chiral marker by adopting the expression of the local Chern marker, Eq.~\eqref{eq:local-chern-marker}, and introducing a one-parameter family of projectors $P_{\vartheta}$ of the form
\begin{equation}
P_{\vartheta}=\frac{1}{2}\left[1-\sin(\vartheta)\left(1-2\rho\right)-\cos(\vartheta)S\right]\label{eq:projector-path-chiral},
\end{equation}
where the parameter $\vartheta$ acts as the extra dimension, resulting in an even dimensional integral over real space.
Here $P_{\pi/2}=\rho$ is the projector of interest and $P_{0}=(1-S)/2$, which is a trivial projector without the same chiral constraint as $\rho$.
Expressed in terms of $P_{\vartheta}$, the local chiral marker is~\cite{Supplemental-Material-arxiv}
\begin{equation}
	\nu(\bm{r})=\!
	\sum_{\alpha}\!\int_0^{\pi/2}\!\!\!\!\!\!\!\!d\vartheta \frac
		{%
			\varepsilon^{i_{0},\dotsc ,i_D}%
			[P_{\vartheta}X_{i_{0}}P_{\vartheta}\dotsb X_{i_{D}}P_{\vartheta}]_{(\mathbf r,\alpha),(\mathbf r,\alpha)}%
		}%
			{[(D+1)/2]!/
			[2i (2\pi i)^{(D-1)/2}]},
			\label{eq:local-chiral-marker}
\end{equation}
where $X_{0}=i\partial_\vartheta$, and $D$ is now odd.
Provided translation invariance, $\nu(\bm{r})\mod 2$ is equal to two times the difference in the Chern-Simons invariant~\cite{nakahara18} between the bundles $\rho$ and $P_{0}$.
In the presence of the chiral constraint, the Chern-Simons invariant is a half integer valued $\mathbb{Z}_2$ invariant~\cite{ryu10}, which implies that $\nu(\bm{r})$ is quantized as an integer.
%
%
%
%
However, by using the $P_\vartheta$ given in Eq.~\eqref{eq:projector-path-chiral} in terms of $S$ and $\rho$, not only $\nu(\bm{r})\mod 2$, but also $\nu(\bm{r})$ is a topological invariant, which in the translation invariant limit equals the chiral winding number~\footnote{ In principle, $\nu(\bm{r})$ and $\nu$ could provide different integers for the same phase.
However, by explicitly verifying that they are equal for examples with $\nu=1$, they are guaranteed to be equal for all phases, which follows from their additive property with respect to direct sums: $\nu_{\rho\otimes\varrho}(\bm{r})=\nu_{\rho}(\bm{r})+\nu_{\varrho}(\bm{r})$}.
Expanding and integrating Eq.~\eqref{eq:local-chiral-marker} the chiral marker becomes~\cite{Supplemental-Material-arxiv}
\begin{equation}
\nu(\mathbf{r})=\gamma_{D}\varepsilon^{i_{1},\dotsc,i_{D}}	\sum_\alpha [\rho SX_{i_{1}}\rho\dotsb \rho X_{i_{D}}\rho]_{({\mathbf{r}\alpha}),(\mathbf{r}\alpha)},\label{eq:local-chiral-winding}
\end{equation}
where $\gamma_{D}$ is a dimension-dependent constant: $\gamma_{1}=-2$, and $\gamma_{3}=-8\pi i/3$. 
By averaging $\nu(\bm{r})$ over a large enough region, it assumes the role of a coarse-grained invariant for non-translationally-invariant systems, in analog to that of the Chern marker.

\textit{The local Chern-Simons marker.}---The topological phases of odd-dimensional systems that break chiral symmetry are defined by the Chern-Simons invariant, a $\mathbb{Z}_2$ invariant which characterizes one of the AZ classes in each odd dimension~\cite{qi08,essin09,ryu10}.
The corresponding real space Chern-Simons marker $\nu_{\rm{cs}}$ is defined almost analogously to the chiral marker in Eq.~\eqref{eq:local-chiral-marker}, but with two crucial differences: since the Chern-Simons marker is a $\mathbb{Z}_2$ invariant, it is only defined modulo 2, and since the chiral symmetry is broken, the path in parameter space must be redefined.
We consider the sum of two paths in parameter space: the first between a trivial state and a chiral state with density matrix $Q$, and the second traversing from the same chiral state to the final state (without a chiral constraint) with density matrix $\rho$.
Constructing the projector for the second path such that it has the same symmetries as the chiral and final states combined, results in a vanishing ($\text{mod }2$) difference in the Chern-Simons marker between the initial and final points in the second path.
This means that the total Chern-Simons marker comes from the path between the trivial and chiral states alone, where again, due to the enforced symmetries, the chiral endpoint gives the same contribution to the marker as $\rho$.

We use the topological insulator in three dimensions (class AII) as a concrete example.
The single particle density matrix of this class obeys the time reversal constraint, $T\rho^*T^{\dagger}=\rho$, where $\mathcal{T}=T\mathcal{K}$ is the time reversal operator in real space and $\mathcal{K}$ is complex conjugation.
We define the single particle density matrix for the chiral state as:
\begin{equation}
Q =\frac{1}{2}\left(1+i|[\rho,S_{R}]|^{-1}[\rho,S_{R}]\right),
\label{eq:Qdef}
\end{equation}
where $|[\rho,S_{R}]|$ is the matrix absolute value~\footnote{The absolute value of a matrix $A$ is defined such that in the eigenbasis of $A$, $|A|$ is diagonal with the absolute values of its eigenvalues in the diagonal.}.
$Q$ obeys a chiral constraint $S_{R}=1-2R$, such that $\{Q,S_{R}\}=S_{R}$.
The operator $R$ can be chosen to be any trivial projector for which $i[\rho,S_{R}] \propto i[\rho,R]$ has no zero modes, and thus renders $Q$ to be a localized operator.
In practice this means that $R$ is model dependent, but as it can be constructed to be any tensor product of local operators we expect that one can always construct $R$ such that the spectrum of $i[\rho,R]$ is gapped~\cite{Supplemental-Material-arxiv}. 
Retaining an $R$ such that $i[\rho,R]$ has zero modes would require fine-tuning.
Provided that both $Q$ and $\rho$ are local operators, any one-parameter family of projectors interpolating between them will be localized as well.
To ensure that the path between the chiral and the final state is confined within the AII class, we demand that the corresponding projector,
\begin{equation}
P_{\vartheta}^{\rho,\rm{Q}}=\frac{1}{2}\left[1-\sin(\vartheta)(1-2\rho)-\cos(\vartheta)(1-2Q)\right],
\label{eq:projector-path-AII}	
\end{equation}
is invariant under time-reversal symmetry, which amounts to enforcing that $TQ^*T^\dagger=Q$. 
Using the definition in Eq.~\eqref{eq:Qdef}, this translates to the condition $TS_{R}^*T^\dagger=-S_{R}$, hence restricting the choice of $R$.
Since the path is contained within the same symmetry class and is always characterized by a localized projector, the corresponding Chern-Simons marker is zero modulo two under this constraint.
The only contribution to the total marker comes from the path between the trivial and chiral state given by the projector
\begin{align}
&P_{\vartheta}^{R}=\frac{1}{2}\left[1-\sin(\vartheta)(1-2Q)-\cos(\vartheta)S_{R}\right]\label{eq:projector-path-chiral2} . 
\end{align}
Since $Q$ obeys a chiral constraint, the total Chern-Simons marker is evaluated by using the local chiral marker in  Eq.~(\ref{eq:local-chiral-winding}) such that $\nu_{\rm{cs}}=\nu\mod 2$.
The approach in this example translates to any odd-dimensional system that breaks chiral symmetry and is characterized by the Chern-Simons invariant; for the one-dimensional superconductor (class D) which obeys the charge-conjugation constraint, one would enforce the projector in Eq.~\eqref{eq:projector-path-AII} to be invariant under charge conjugation.

\begin{figure*}[t]
  \includegraphics[width=1\textwidth]{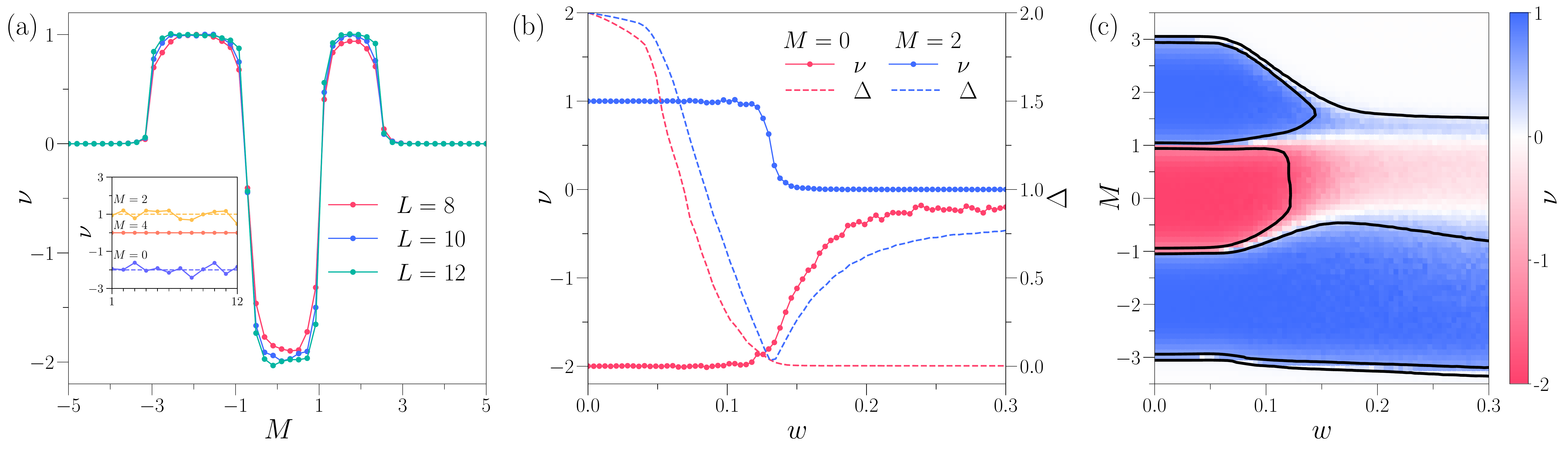}
    \caption{Chiral markers for a topological superconductor Hamiltonian Eq.~\eqref{eq:Hamiltonian} with $t=0$ and $\lambda=1$. (a) The chiral marker $\nu$ as a function of the mass parameter, $M$, for three different system sizes: $L=8$, $L=10$, and $L=12$. The inset shows the (nonaveraged) value of $\nu$ for $M=0$, $M=2$, and $M=4$ at $12$ randomly chosen sites in the lattice. (b) The chiral marker (left $y$ axis), and the energy gap, $\Delta$, (right $y$ axis) as a function of the Gaussian width $w$ for $M=0$ and $M=2$ for system size $L=12$. (c) The phase diagram in the parameter space $(M, w)$. The superimposed curve represents the opening or closing of the energy gap, where as a reference, the gap is open at the point $M=0, w=0$. In (a, c) $\nu$ is averaged over 25 sites and four lattice realizations, while in (b) we considered 10 lattice realizations. In (b),(c) $\Delta$ is averaged over 100 lattice realizations.}
   \label{fig:DIII}
\end{figure*}   

\textit{Application to topological amorphous solids.}---
In this section we provide examples that demonstrate how the chiral and Chern-Simons markers can be used to numerically characterize topological phases in amorphous systems.
We consider two three-dimensional models, a topological insulator (class AII) and a topological superconductor (class DIII), both with the same first quantized Hamiltonian: 
\begin{gather}
    \label{eq:Hamiltonian}
    \mathcal H_{ij} = -\delta_{ij}M \tau_z-2t_{ij}\tau_z +\\ t_{ij}
    \big\lbrace [\sigma_z\cos \theta + e^{-i\phi} (\sigma_x + i \sigma_y)\sin \theta\ ][i\lambda\tau_x+t \tau_y ]+\text{H.c.}\big\rbrace \notag,
\end{gather}
with $\hbar=1$.
Here $\phi$ and $\theta$ are the azimutal and polar angles between lattice sites $i$ and $j$, $t$ and $t_{ij}=1/4\exp(1-\vert \pos_i - \pos_j \vert /a)$ are hopping amplitudes \cite{Agarwala2017} (with $a$ the average bond length), $M$ is a mass parameter, $\sigma_\alpha$ and $\tau_\alpha$ with $\alpha\in(x,y,z)$ are Pauli matrices where the $\sigma_\alpha$ acts on the spin degrees of freedom. 
For the superconductor $\lambda$ represents a pairing potential, and $\tau_\alpha$ acts in particle-hole space, while for the insulator, $\lambda$ is a spin-orbit coupling parameter, and $\tau_\alpha$ acts in orbital space. 
For $\theta\in\{0, \pi\}$ and $\phi\in\{0, \pm \pi/2, \pi\}$, $\mathcal H_{ij}$ restricts to a cubic lattice Hamiltonian, known to host different topological phases depending on the parameters~\cite{wang15}.
To make the lattice amorphous we choose the position of each lattice site from a Gaussian distribution with a standard deviation $w$, centered on the lattice site positions of the crystalline limit.
One can therefore continuously tune the lattice from a crystalline structure when $w=0$ to an increasingly amorphous one as $w$ increases.
The model has a fixed number of six nearest neighbors which are not necessarily the same as in the crystalline limit, allowing for defects to enter as $w$ increases.

The time-reversal invariant superconductor in class DIII obeys a chiral constraint and its phases are characterized using the chiral marker Eq.~\eqref{eq:local-chiral-winding}. 
In particular, setting $t=0$, the chiral constraint for the Hamiltonian in Eq.~\eqref{eq:Hamiltonian} is imposed by $S=-\tau_y$.
The averaged chiral marker for the amorphous lattice with width $w=0.1$ is depicted in Figure~\ref{fig:DIII}(a) as a function of $M$ for three different system sizes, $L=8$, $L=10,$ and $L=12$.
As the system size increases, the chiral marker approaches an integer-quantized value, as expected in the absence of finite size effects.
In the crystalline limit the chiral marker can be evaluated analytically and takes three different values depending on the parameter $M$, $\nu = -2$ for $|M|< 1$, $\nu = 1$ for $1<|M|< 3$ and $\nu=0$ for all other values of $M$~\cite{ryu10}.
The inset in Figure~\ref{fig:DIII}(a) shows the chiral marker of twelve randomly selected sites of a single realization of the lattice for $M=0,2,4$, to highlight how the the marker fluctuates around its mean value.
The effect of the amorphicity is apparent in the topological transition from $\nu=1$ to $\nu=0$ at $M\simeq 2.3$, which is slightly displaced compared to the crystalline limit.
Figure~\ref{fig:DIII}(b) shows the chiral marker as a function of $w$ for fixed $M$. 
For $M=2$ the gap closes only at a single point and there is a conventional topological phase transition to a trivial state, but driven by amorphicity.
This type of transition has also been reported in~\cite{Wang2022}.
For $M=0$, on the other hand, the energy gap remains closed for increasing $w$ after the transition.
Figure~\ref{fig:DIII}(c) is the complete phase diagram in the parameter space $(M, w)$.
The black line indicates the regions where the gap crosses a certain energy threshold of $E^*=0.12$.
The different type of transitions discussed above are distinguished by the presence of either one single line, indicating that the gap closes and does not reopen, or two lines, indicating where the gap closes and reopens, along the transition line between phases.
The lower part of the phase diagram shows that the topological phase $\nu=1$ survives deep into the amorphous regime, exemplifying the fact that the topological properties of the crystalline case can survive, or even be enhanced, by the process of amorphization.
\begin{figure}[b]
  \includegraphics[width=1\linewidth]{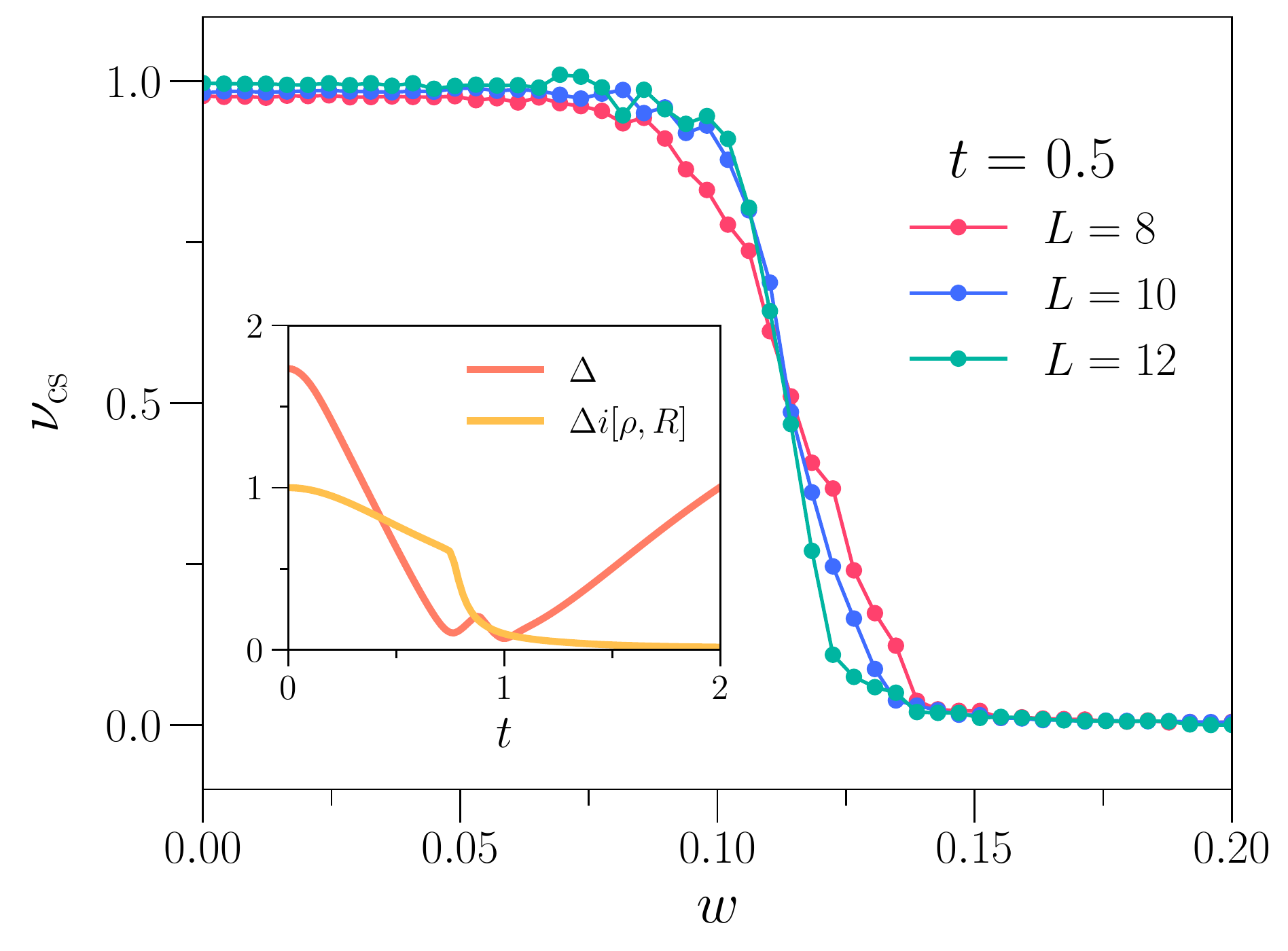}
    \caption{Chern-Simons marker for a topological insulator Hamiltonian Eq.~\eqref{eq:Hamiltonian} with $\lambda=1$. The local Chern-Simons marker, $\nu_{\rm{cs}}$ as a function of $w$, for system sizes: $L=8$, $L=10$, $L=12$. The inset shows the spectral gap, $\Delta$, and the gap of the operator $i[\rho, R]$ as a function of $t$, for system size $L=12$, $w=0.05$ and $M=2$. In the main figure $\nu_{\rm{cs}}$ is averaged over 25 sites and four lattice realizations.}
   \label{fig:AII}
\end{figure} 

For the three-dimensional topological insulator in class AII there is no chiral constraint and one needs to choose the operator $R$~\cite{Supplemental-Material-arxiv}.
A natural choice is $R = (1-S)/2$, with $S$ the chiral constraint of the Hamiltonian in Eq.~\eqref{eq:Hamiltonian} in the limit where $t=0$.
As shown in the inset of Figure~\ref{fig:AII} the operator $i[\rho,R]$ is gapped within the range $0\leq t\lessapprox 1$ for any width $w$ we consider, thus satisfying the necessary conditions in order for $Q$ to be a localized~\footnote{Exploring different parameter ranges will require a different choice of $R$ built as a tensor product of multiple-site operators.}.
In Figure~\ref{fig:AII} we show the Chern-Simons marker as a function of $w$ for fixed $t=0.5$ and $M=2$, for different system sizes.
For these parameters the system is in the topological phase $\nu_{\rm{cs}}=1$ in the crystalline limit.
This phase survives in the amorphous regime until $w\simeq0.1$.

{\it Discussion.}--- We have provided a general expression for local topological markers for topological phases in odd dimensions characterized by the chiral winding number and the Chern-Simons invariant.
We derived this expression by considering an extra dimension in the form of a one-parameter family of projectors between an atomic limit state and a non-trivial topological state, which allowed us to reformulate the chiral and Chern-Simons invariants in the form of a Chern marker.
We have confirmed the validity of these local topological markers by using them to characterize the topological phases of amorphous superconductors and insulators in three dimensions.
Although we have only considered Slater determinants, our expressions for the markers apply more generally to all states where the spectrum of $\rho$ is gapped; this is typically the case for interacting localized states, see, e.g., Ref.~\cite{jens15}.
With such a gap, one can define a local band-flattened single-particle density matrix, $\varrho=[1+|2\rho-1|^{-1}(2\rho-1)]/2$, and the result for Slater determinants carries over to the interacting case by replacing $\rho$ by $\varrho$.

We are grateful to A. Tiwari for fruitful discussions and input in the early stages of this work. 
We also thank D. Aceituno for the help with various numerical aspects.
The research of T.K.K is funded by the Wenner-Gren Foundations.
This work received funding from the European Research Council (ERC) under the European Union’s Horizon 2020 research and innovation program (Grant Agreement No. 101001902),
and from the Swedish Research Council (VR) through Grants No. 2019-04736 and No. 2020-00214.

\bibliography{refs}
\newpage

\onecolumngrid

\setcounter{secnumdepth}{5}
\renewcommand{\theparagraph}{\bf \thesubsubsection.\arabic{paragraph}}

\renewcommand{\thefigure}{SM\arabic{figure}}
\setcounter{figure}{0} 
\appendix

\section*{Supplemental Material}

\section{A derivation of the Chiral marker}

We present the Chern number as the integral over the Brillouin zone of the Chern character, expressed in terms of the single particle density matrix and describe how to express the Chern marker by Fourier transforming the Chern character. 
The same prescription applies to the Chern-Simons invariant, from which we derive the chiral marker.

\subsection{The Chern marker: Fourier-transforming the Chern character}
The Bogoliubov-de Gennes single-particle density matrix for a many-body state $\ket{\phi}$ is a matrix acting in single-particle space, with indices 
\begin{align}
	\rho_{\alpha\beta}=\braket{\phi|\psi_{\beta}^{\dagger}\psi_{\alpha}|\phi},
\end{align}
where $\{\psi_{\alpha}^{\dagger}\}$ are Bogoliubov-de Gennes creation operators, where, $s$ is the particle hole index ($s = \pm 1 $) and $a$ labels the single-particle space, $\psi_{a,s}^{\dagger}=\delta_{s,1}c_{a}^{\dagger}+\delta_{s,-1}c_{a}$, where $c^\dagger$ and $c$ are the usual creation and annihilation operators. 
If $\ket{\phi}$ has a unitary symmetry then $\rho$ has a block-diagonal structure. 
For example, for states with a $U(1)$ fermion number conservation symmetry (states with a definite fermion number) $\rho$ takes the block-diagonal form,
\begin{equation}
\rho=\begin{pmatrix}\rho^{U(1)} & 0\\
0 & 1-(\rho^{U(1)})^{*}
\end{pmatrix},\label{app:eq:block-diagonal}
\end{equation}
where $\rho^{U(1)}$ is the matrix with elements $\rho_{ab}^{U(1)}=\braket{\phi|c_{b}^{\dagger}c_{a}|\phi}$.
It is these blocks that enter the topological classification, and in the following discussion we refer to them as $\rho$, or the single particle density matrix.

In terms of the single-particle density matrix, the n'th Chern character~\cite{nakahara18} takes the form
\begin{equation}
ch_n = \frac{1}{(2\pi i)^{n}}\frac{1}{n!}\varepsilon^{i_1,\dotsc , i_{2n}}\text{Tr}\left(\tilde\rho(\mathbf k)\partial_{k_{i_1}}\tilde\rho(\mathbf k)\dotsb \tilde\rho(\mathbf k)\partial_{k_{i_{2n}}}\tilde\rho(\mathbf k)\right),
\end{equation}
where the lattice spacing is set to unity, $\varepsilon$ is the Levi-Civita symbol, repeated indices are summed over, and $\tilde\rho(\mathbf{k})$ are matrices in the local Hilbert space, they are the Fourier components of the single particle density matrix $\rho$, 
\begin{equation}
\rho=(2\pi)^{D}\int_{BZ}d^{D}k\,\tilde\rho(\mathbf{k})\ket{\mathbf{k}}\bra{\mathbf{k}}.
\end{equation}
Here and in the rest of these notes kets, $\ket{\cdot}$, refer to single-particle states.
Integrating the Chern character over an even dimensional ($D=2n$) surface results in a Chern number characterizing the Brillouin zone vector bundle.
Taking the surface to be the first Brillouin zone, $BZ$,  yields the Chern number usually referred to when discussing free fermion topological states:
\begin{equation}
\mathcal{C} = \frac{1}{(2\pi i)^{D/2}}\frac{1}{(D/2)!}\int_{BZ}d^{D}k\varepsilon^{i_1,\dotsc ,i_{D}}\text{Tr}\left(\tilde\rho(\mathbf k)\partial_{k_{i_1}}\tilde\rho(\mathbf k)\dotsb \tilde\rho(\mathbf k)\partial_{k_{i_D}}\tilde\rho(\mathbf k)\right).
\end{equation}

Acting with $\rho$ on a state 
\begin{equation}
\ket{\psi}=\frac{1}{(2\pi)^{D}}\int_{BZ}d^{D}k\tilde{\psi}(\mathbf{k})\ket{\mathbf{k}},
\end{equation}
where $\tilde{\psi}(\mathbf{k})$ is a vector in the local Hilbert space, yields
\begin{equation}
\rho\ket{\psi}=\frac{1}{(2\pi)^{D}}\int_{BZ}d^{D}k\tilde\rho(\mathbf{k})\tilde{\psi}(\mathbf{k})\ket{\mathbf{k}},
\end{equation}
since $\braket{\mathbf{k}^{\prime}|\mathbf{k}}=(2\pi)^{-D}\delta^{D}(\mathbf{k}-\mathbf{k}^{\prime})$.
By defining the eigenvalue of the position operator $X_i$ through $X_i\ket{\mathbf{r}}=x_i\ket{\mathbf{r}}$ and using the Fourier decomposition of the momentum vector $\ket{\mathbf{k}}=\sum_{\mathbf{r}}e^{-i\mathbf{k}\cdot\mathbf{r}}\ket{\mathbf{r}}$, the action of the position operator on the state $\ket{\psi}$ evaluates as
\begin{equation}
\begin{aligned}
X_{i}\ket{\psi}&=\int_{BZ}d^{D}k\tilde{\psi}(\mathbf{k})X_{i}\ket{\mathbf{k}}\\
&=\sum_{\mathbf{r}}\int_{BZ}d^{D}k\tilde{\psi}(\mathbf{k})e^{-i\mathbf{k}\cdot\mathbf{r}}x_{i}\ket{\mathbf{r}}\\
&=\sum_{\mathbf{r}}\int_{BZ}d^{D}k\tilde{\psi}(\mathbf{k})i\partial_{k_{i}}e^{-i\mathbf{k}\cdot\mathbf{r}}\ket{\mathbf{r}}\overbrace{=}^{\text{partial integration}}-i\sum_{\mathbf{r}}\int_{BZ}d^{D}k(\partial_{k_{i}}\tilde{\psi}(\mathbf{k}))e^{-i\mathbf{k}\cdot\mathbf{r}}\ket{\mathbf{r}}\\
&=-i\int_{BZ}d^{D}k(\partial_{k_{i}}\tilde{\psi}(\mathbf{k}))\ket{\mathbf{k}}.
\end{aligned}
\end{equation}
A localized wave function is up to a phase a constant vector $\alpha$ in the local Hilbert space, $\tilde{\psi}(\mathbf{k})=\alpha e^{i\mathbf{k}\cdot\mathbf{r}}$, and by denoting the corresponding localized state vector as: $\ket{\psi}=\ket{\mathbf{r};\alpha}$, the expectation value $\braket{\mathbf{r};\alpha|\rho X_{i}\rho X_{j}\dotsb\rho|\mathbf{r};\alpha}$ becomes:
\begin{equation}
\begin{aligned}
\label{app:eq:fourier_exp}
\braket{\mathbf{r};\alpha|\rho X_{i_1}\rho X_{i_2}\dotsb  X_{i_{D}}\rho|\mathbf{r};\alpha}&=(-i)^{D}\int_{BZ}d^{D}k^{\prime}\int_{BZ}d^{D}k\alpha^{\dagger}\tilde\rho(\mathbf{k})\partial_{k_{i_1}}\tilde\rho(\mathbf{k})\partial_{k_{i_1}}\dotsb \partial_{k_{i_{D}}}\tilde\rho(\mathbf{k})\alpha e^{i(\mathbf{k-k^{\prime})}\cdot\mathbf{r}}\braket{\mathbf{k}^{\prime}|\mathbf{k}}\\
&=(2\pi i)^{-D}\int_{BZ} d^{D}k\,[\tilde\rho(\mathbf{k})\partial_{k_{i_1}}\tilde\rho(\mathbf{k})\partial_{k_{i_2}}\dotsb \partial_{k_{i_{D}}}\tilde\rho(\mathbf{k})]_{\alpha,\alpha}.
\end{aligned}
\end{equation}
Using Eq.~\eqref{app:eq:fourier_exp}, the Chern number is given by
\begin{equation}
\begin{aligned}
\mathcal{C}&=\frac{1}{(2\pi i)^{D/2}}\frac{1}{(D/2)!}\int_{BZ}d^{D}k\varepsilon^{i_1,\dotsc, i_{D}}\text{Tr}\left(\tilde\rho(\mathbf k)\partial_{k_{i_1}}\tilde\rho(\mathbf k)\dotsb \tilde\rho(\mathbf k)\partial_{k_{i_D}}\tilde\rho(\mathbf k)\right)\\
&=\frac{(2\pi i)^{D/2}}{(D/2)!}\sum_{\alpha}\varepsilon^{i_1,\dotsc ,i_{D}}\braket{\mathbf{r};\alpha|\rho X_{i_1}\rho X_{i_2}\dotsb  X_{i_{D}}\rho|\mathbf{r};\alpha}.
\end{aligned}
\end{equation}
The last expression is the local Chern marker, which we in the main article present as 
\begin{equation}
	\mathcal C(\mathbf{r})=\sum_{\alpha}\frac{\varepsilon^{i_{1},\dotsc,i_{D}} [\rho  	X_{i_{1}}\rho X_{i_{2}}\dotsb X_{i_{D}}\rho]_{(\mathbf r,\alpha),(\mathbf r,\alpha)}}{(D/2)!/(2\pi i)^{D/2}},
\label{app:eq:local-chern-marker}
\end{equation}
where the expectation value is expressed as $\braket{\mathbf{r};\alpha|\rho X_{i_1}\rho X_{i_2}\dotsb  X_{i_{D}}\rho|\mathbf{r};\alpha}=[\rho X_{i_{1}}\rho X_{i_{2}}\dotsb X_{i_{D}}\rho]_{(\mathbf r,\alpha),(\mathbf r,\alpha)}$.

\subsection{The Chern-Simons invariant and the local chiral marker}

The Chern number is a bundle invariant in even dimensions alone.
The corresponding bundle invariant for odd dimensions is the Chern-Simons invariant $\mathcal{CS}$, which is the integral of the Chern-Simons form $cs_n$ over an odd dimensional surface~\cite{nakahara18}.
The Chern-Simons invariant is not a local marker as it cannot be expressed as a sum of local expectation values---it is the modulo one part of a basis dependent expression and cannot be expressed in terms of the single particle density matrix alone.
However, the Chern character and the Chern-Simons form are connected in that the integrals of the Chern-Simons forms over closed $D-1$-dimensional surfaces $\partial \Omega\subset \Omega$ equals integrals over Chern characters over the even $D$-dimensional surface $\Omega$:
\begin{equation}
\int_{\partial \Omega}cs_n=\int_\Omega ch_n.
\end{equation}
By introducing a one dimensional parameter family of projectors $P_\vartheta$ such that $P_0$ is a trivial projecor and $P_{\pi/2}=\rho$ is the projector of interest, the difference in the Chern-Simons invariant, $\Delta \mathcal{CS}$ between these two states becomes
\begin{equation}
\Delta \mathcal{CS}=\int_{BZ} d^{D}k\, cs_n=\int_0^{\pi/2}\diff \vartheta \int_{BZ} d^{D}k\, ch_n,
\end{equation}
where the Brillouin zone is odd dimensional and the Chern character is:
\begin{equation}
ch_n = \frac{1}{(2\pi i)^{(D+1)/2}}\frac{1}{((D+1)/2)!}\varepsilon^{\vartheta,i_1,\dotsc , i_D}\text{Tr}[\mathcal{P}(\mathbf{k})\partial_{\vartheta}\mathcal{P}(\mathbf{k})\partial_{k_{i_1}}\dotsb  \mathcal{P}(\mathbf{k})\partial_{k_{i_D}}\mathcal{P}(\mathbf{k})],
\end{equation}
where $\mathcal{P}$ denotes the Fourier component of $P$.
So, introducing the extra dimension through the parameter $\vartheta$ allows us to express the Chern-Simons invariant as the integral of the Chern character over an even dimensional surface, which in turn is a well defined expression in terms of the projector.

Using Eq.~(\ref{app:eq:fourier_exp}),
\begin{equation}
\braket{\mathbf{r};\alpha|Pi\partial_{\vartheta}PX_{i_1}\dotsb PX_{i_D} P|\mathbf{r};\alpha}=i(2\pi i)^{-D}\int d^{D}k\,[\mathcal{P}(\mathbf{k})\partial_{\vartheta}\mathcal{P}(\mathbf{k})\partial_{k_{i_1}}\dotsb  \mathcal{P}(\mathbf{k})\partial_{k_{i_D}}\mathcal{P}(\mathbf{k})]_{\alpha,\alpha}.
\end{equation}
The difference in Chern-Simons invariants is therefore
\begin{equation}
\begin{aligned}
\Delta CS&=\frac{1}{(2\pi i)^{(D+1)/2}}\frac{1}{((D+1)/2)!}\sum_{\alpha}\int d\vartheta\int_{BZ}d^{D}k\varepsilon^{\vartheta,i_1,\dotsc,i_D }[\mathcal{P}(\mathbf{k})\partial_{\vartheta}\mathcal{P}(\mathbf{k})\partial_{k_{i_1}}\dotsb  \mathcal{P}(\mathbf{k})\partial_{k_{i_D}}\mathcal{P}(\mathbf{k})]_{\alpha,\alpha}\\
&=\overbrace{\frac{i(2\pi i)^{D}}{(2\pi i)^{(D+1)/2}}}^{i(2\pi i)^{(D-1)/2}}\frac{1}{((D+1)/2)!}\sum_{\alpha}\int d\vartheta\int_{BZ}d^{D}k\varepsilon^{\vartheta,i_1,\dotsc,i_D}\braket{\mathbf{r};\alpha|Pi\partial_{\vartheta}PX_{i_1}\dotsb PX_{i_D} P|\mathbf{r};\alpha}.
\end{aligned}
\end{equation}

A projector $\rho$ with a chiral constraint defines the local chiral marker as $\nu(\bold{r}) \mod 2=2 \Delta CS$, which is integer valued since the Chern-Simons invariant in the presence of a chiral constraint is a half integer valued $\mathbb{Z}_2$ invariant.

The chiral marker as presented in the main text, is thus,
\begin{equation}
	\nu(\bm{r})=
	2i\sum_{\alpha}\int_0^{\pi/2}\!\!\!\!\!\!\!\!d\vartheta \frac
		{%
			\varepsilon^{i_{0},\dotsc,i_{D}}%
			[P_{\vartheta}X_{i_{0}}P_{\vartheta}\dotsb X_{i_{D}}P_{\vartheta}]_{(\mathbf r,\alpha),(\mathbf r,\alpha)}%
		}%
			{[(D+1)/2]!/(2\pi i)^{(D-1)/2}},
			\label{app:eq:local-chiral-marker}
\end{equation}
where $X_0=i\partial_\vartheta$.

In Eq.~\eqref{app:eq:local-chiral-marker}, one may use any localized interpolation between a trivial state and $\rho$ to get an integer related to the Chern-Simons invariant.
However, different choices for $P$ have different finite-size corrections to exact quantized integer values.
For a finite system of linear size $L$ these corrections scale as  $\sim e^{-\xi/L}$, where $\xi$ is the localization length of $P$. 
Since $P$ equals the single-particle density matrix in question, $\rho$, for one value of $\vartheta$, it cannot be more localized than $\rho$. 
The choice 
\begin{equation}
P_{\vartheta}=\frac{1}{2}\left[1-\sin(\vartheta)\left(1-2\rho\right)-\cos(\vartheta)S\right]\label{app:eq:projector-path-chiral},
\end{equation}
where $S$ is the chiral constraint $\{S,\rho\}=S$, has the same localization length as $\rho$ and is, therefore, the best one can do.
Expanding Eq.~\eqref{app:eq:local-chiral-marker} with this expressions and using that products of the position operators $X_i X_j$ vanishes when contracted with the Levi-Civita symbol, and integrating over $\vartheta$ leaves only terms with either a $\rho$, or a $S$, or their product, in between two $X_i$'s:
\begin{align}
\sum_{\alpha}\varepsilon^{i_{1},\dotsc,i_{D}}\braket{\mathbf r,\alpha|\rho X_{i_{1}}\rho\dotsb X_{i_{D}}\rho|\mathbf r,\alpha}\label{app:eq:term1}\\
\sum_{\alpha}\varepsilon^{i_{1},\dotsc,i_{D}}\braket{\mathbf r,\alpha|\rho S X_{i_{1}}\rho\dotsb X_{i_{D}}\rho|\mathbf r,\alpha}\label{app:eq:term2}\\
\sum_{\alpha}\varepsilon^{i_{1},\dotsc,i_{D}}\braket{\mathbf r,\alpha|\rho  X_{i_{1}}\rho S  X_{i_{2}} \rho\dotsb X_{i_{D}}\rho|\mathbf r,\alpha}\label{app:eq:term3}\\
\text{etc}.
\end{align}
By using the relations $\{S,\rho\}=S$ and $[S,X_{i}]=0$ we can move the $S$'s around and place them all at the same place, which by using ${S}^{2}=1$ leaves only two non-equivalent terms: the terms in Eq.~(\ref{app:eq:term1}) and Eq.~(\ref{app:eq:term2}) above. 
However, the term in Eq.~(\ref{app:eq:term1}) vanishes in the translation invariant limit. 
Using the anti-symmetry of the Levi-Civita symbol to insert commutators into Eq.~(\ref{app:eq:term1}) gives
\begin{equation}
\begin{aligned}
\sum_{\alpha}\braket{\alpha,\mathbf{r}|\varepsilon^{i_{1},\dotsc ,i_{D}}\rho[X_{i_{1}},\rho]\dotsb[X_{D},\rho]|\mathbf{r},\alpha}&
\propto\varepsilon^{i_{1},\dotsc ,i_{D}}\int_{BZ} d^{D}\text{Tr}(\tilde\rho(\mathbf k)[X_{i_{1}},\tilde\rho(\mathbf k)]\dotsb[X_{2n+1},\tilde\rho(\mathbf k)])\\
&=\varepsilon^{i_{1},\dotsc ,i_{D}}\int_{BZ} d^{D}\text{Tr}(\tilde\rho(\mathbf k)[X_{i_{1}},\tilde\rho(\mathbf k)]\dotsb[X_{D},\tilde\rho(\mathbf k)]\tilde\rho(\mathbf k),
\end{aligned}
\end{equation}
which vanishes, since for any string of operators 
\begin{equation}
Q[A_{1},Q][A_{2},Q]\dotsb[A_{D},Q]Q=0,
\label{app:eq.operatorsstring}
\end{equation}
where $Q$ is a projection operator and the $\{A_{i}\}_{i}$ are arbitrary operators and $D$ is an odd integer. 
This statement can be shown by induction.
The final form of the chiral marker is therefore
\begin{equation}
\nu(\mathbf{r})=\gamma_{D}\varepsilon^{i_{1},\dotsc,i_{D}}	\sum_\alpha [\rho SX_{i_{1}}\rho\dotsb \rho X_{i_{D}}\rho]_{({\mathbf{r}\alpha}),(\mathbf{r}\alpha)},\label{app:eq:local-chiral-winding}
\end{equation}
where $\gamma_{D}$ is a dimension dependent constant.

\section{Explicitly expanding and simplifying the expression for the chiral marker in three spatial dimensions}

We expand the general expression of the chiral marker in three spatial dimensions and describe how to implement it in practice when considering the time reversal invariant superconductor on an amorphous lattice.

\subsection{Simplifying the expression for the local marker}
The local chiral marker, Eq.~\eqref{app:eq:local-chiral-marker}, in three spatial dimensions is
\begin{equation}
	\nu(\bm{r})=
	-2\pi \sum_{\alpha}\int_0^{\pi/2}d\vartheta \hspace{1mm}
			\varepsilon^{\mu\nu\rho\sigma}
			[P_{\vartheta}X_{\mu}P_{\vartheta}X_{\nu}P_{\vartheta}X_{\rho}P_{\vartheta}X_{\sigma}P_{\vartheta}]_{(\mathbf r,\alpha),(\mathbf r,\alpha)}%
	.
			\label{app:eq:local-chiral-marker3d}
\end{equation}
where Greek letters, $\mu\in(0,1,2,3)$.
$X_{0}=i\partial_\vartheta$, and $X_{i}$, $i\in(1,2,3)$ is the i'th position operator: $(X_i)_{(\mathbf r,\alpha),(\mathbf r^\prime,\beta)}=x_i \delta_{\alpha,\beta}\delta_{\mathbf r,\mathbf r^\prime}$, where $x_i$ is the $i$'th component of the position $\mathbf r$, and $\alpha,\beta$ denote local quantum numbers.

$P_\vartheta$ is the family of projectors,
\begin{equation}
\label{app:eq:projector-path-chiral}
P_{\vartheta}=\frac{1}{2}\left[1-\sin(\vartheta)\left(1-2\rho\right)-\cos(\vartheta)S\right].
\end{equation}
where $\rho$ is the single particle density matrix of interest, the and the operator $S$ obeys the chiral constraint, $\{\rho,S\}=S$.
For notational convenience we will from now on refer to $P_\vartheta$ as $P$.
Expanding the operator in Eq.~\eqref{app:eq:local-chiral-marker3d}, and using the anti-symmetry of the Levi-Civita symbol, yields the expression
\begin{equation}
\begin{aligned}
\varepsilon^{\mu\nu\rho\sigma}
			PX_{\mu}PX_{\nu}PX_{\rho}PX_{\sigma}P&=i\varepsilon^{ijk}\left(P \partial_\vartheta PX_i P X_j PX_k P\right. \\
			&\left.-P X_i P \partial_\vartheta P X_j P X_k P+P X_i P X_j P \partial_\vartheta P X_k P-P X_i P X_j P X_k P \partial_\vartheta P\right),
\end{aligned}
\end{equation}
which is simplified by evaluating the derivatives and using that $(1-2\rho)=S^2=1$, that the chiral and position operators commute, $[S,X_i]=0$, and the chiral constraint, $\{\rho,S\}=S$ to move the $S$'s around.
The translational invariance of the operator in space allows us to throw away any terms with a position operator to the far right or left, as these vanish when taking the expectation value.
The $\vartheta$ dependence of each term factorizes, and after integrating over $\vartheta$ we are left with:
\begin{equation}
\begin{aligned}
-2\pi\sum_\alpha\int_0^{\pi/2}\diff \vartheta\varepsilon^{\mu\nu\rho\sigma}
			[PX_{\mu}PX_{\nu}PX_{\rho}PX_{\sigma}P]_{(\mathbf r,\alpha),(\mathbf r,\alpha)}=-2\pi\sum_\alpha\varepsilon^{ijk}\Big[\big(\frac{4i}{3} \rho SX_i\rho X_j\rho X_k \rho +\frac{i}{2} \rho X_i \rho X_j\rho X_k \rho\big)\Big]_{(\mathbf r,\alpha),(\mathbf r,\alpha)}
\end{aligned}
\end{equation}
The second term is zero which can be seen by using the translational invariance of the operator, and the identity in Eq.~\eqref{app:eq.operatorsstring} to express this term in terms of commutators:
\begin{equation}
\begin{aligned}
\varepsilon^{ijk}
\sum_\alpha\left[\rho X_i \rho X_j\rho X_k \rho\right]_{(\mathbf r,\alpha),(\mathbf r,\alpha)}=&\varepsilon^{ijk} \sum_\alpha\left[\rho [X_i ,\rho ][X_j,\rho ][X_k, \rho]\right]_{(\mathbf r,\alpha),(\mathbf r,\alpha)}\\
=&\varepsilon^{ijk} \sum_\alpha\left[\rho [X_i ,\rho ][X_j,\rho ][X_k, \rho]\rho\right]_{(\mathbf r,\alpha),(\mathbf r,\alpha)}\\
=&0
\end{aligned}
\end{equation}
where in the second step the projector to the far left is written as $\rho=\rho^2$, where one of these two projectors is moved to the far right by using the cyclic property of the trace.
Therefore the chiral marker in three dimensions takes the form:
\begin{equation}
\begin{aligned}
\label{app:eq:marker3D-final}
	\nu(\bm{r})=-\frac{8\pi i}{3}\sum_\alpha\varepsilon^{ijk}[ \rho SX_i\rho X_j\rho X_k \rho ]_{(\mathbf r,\alpha),(\mathbf r,\alpha)}.
\end{aligned}
\end{equation}

\subsection{Numerical implementation of the chiral marker}

Here we describe how the local marker is evaluated in practice, by using the amorphous topological superconductor in three dimensions as an example.
The first quantized Hamiltonian describing this model is
\begin{equation}
    \label{app:eq:Hamiltonian}
     H_{ij} = -\delta_{ij}M \tau_z-2t_{ij}\tau_z + t_{ij}
    \big\lbrace [\sigma_z\cos \theta + e^{-i\phi} (\sigma_x + i \sigma_y)\sin \theta\ ][i\lambda\tau_x+t \tau_y ]+h.c.\big\rbrace,
\end{equation}
where $\hbar=1$, $\phi$ and $\theta$ are the azimutal and polar angles between lattice sites $i$ and $j$, and $t$ and $t_{ij}=1/4\exp(1-\vert \pos_i - \pos_j \vert /a)$ are hopping amplitudes, $a$ being the average bond length~\cite{Agarwala2017}.
$M$ is a mass parameter, $\lambda$ represents a pairing potential, and $\sigma_\alpha$ and $\tau_\alpha$ with $\alpha\in(x,y,z)$ are Pauli matrices, with the former acting on the spin degrees of freedom, and the latter in particle-hole space.

Evaluating the chiral marker $\nu(\pos)$ in Eq.~\eqref{app:eq:marker3D-final} requires explicit expressions for the single particle density matrix $\rho$, the matrix $S$ and the set of position operators $X_i$. 
These operators are $4N$-dimensional square matrices where $N$ is the number of sites, and the factor of four is the orbital degree of freedom.
By choosing the basis given by $\lbrace\vert \pos_i \rangle \otimes \vert \alpha_j \rangle \rbrace$, where $\lbrace\vert \pos_i\rangle\rbrace_{i=1,...,N}$ is the basis of states localized on the amorphous lattice sites and $\lbrace\vert \alpha_j \rangle\rbrace_{j=1,2,3,4}$ is the basis of local quantum numbers, the relevant operators are expressed as:
\begin{align}
    & H = \sum_{ij} H_{ij}\ket{\mathbf r_i}\bra{\mathbf r_j},\\
   &S= -\mathds{1}_N \otimes \mathds{1}_2 \otimes \tau_y,\\
   & X_i = \text{diag}_N(x_i)\otimes\mathds{1}_2 \otimes \mathds{1}_2,
   \label{app:eq:Xi}
\end{align}
where the position matrix $X_i$ includes the diagonal matrix $\text{diag}_N(x_i)$, where $x_i$ is a vector containing the value of the coordinate $i$ in every lattice site.
The matrix $\rho$ is the projector onto all occupied bands, and is constructed from the eigenvectors $\phi_i $ of the Hamiltonian matrix $H$,
\begin{align}
	\rho = \sum_{i\in \rm{occ}}\phi_i   \phi_i^\dagger.
\end{align} 

The form of the position operator in Eq.~(\ref{app:eq:Xi}) works well when considering open boundary conditions.
However, for closed boundary conditions the position operators are not globally well defined.
In Eq.~(\ref{app:eq:Xi})  this fact expresses itself in the jump that the $x_i$ vector takes between the last lattice site, $x_i=N$, and the first one, $x_i=0$, even though these are neighboring sites.
Nevertheless, there exist locally well defined position operators in any simply connected region, any region that does not wrap around the entire system.
When calculating the local marker at $\pos$,  one should use the position operator that is well defined in a region around $\pos$ that is as large as possible, which amounts to defining the position operator as far from the branch cut as possible.
In this case, since the operators involved in the computation of the marker have a certain localization length, the effect of the branchut is minimized by being furthest away as possible from the site where we perform the calculation.
Therefore, to compute the marker we define the matrix $X_i$ for each site $i$ such that the site at position $\pos$ is shifted to be in the center of the system, which allows for the marker at each site to be evaluated as far from the branchcut as possible.
The set of shifted vectors $x'_i$ are defined as
\begin{align}
   & x'_i=\text{mod}(x_i + \delta_{x_i}, N_i )\\
    &\delta_{x_i} = \Theta(x_i -  \lfloor N_i/2 \rfloor )\vert x_i - (N_i +  \lfloor N_i/2 \rfloor ) \vert + \Theta( \lfloor N_i/2 \rfloor  - x_i)\vert  \lfloor N_i/2 \rfloor  - x_i \vert,
\end{align}
where $N_i$ is the number of lattice sites in direction $i$, and $\lfloor N_i/2 \rfloor$ is taken to be the lowest integer closer or equal to $N_i/2$.

The chiral marker is evaluated by using the matrices $\rho$, $S$, and $X_i$ to calculate the operator in Eq.~\eqref{app:eq:marker3D-final}, taking the expectation value and tracing over the orbital degrees of freedom.

\begin{figure}[b]
  \includegraphics[width=0.5\linewidth]{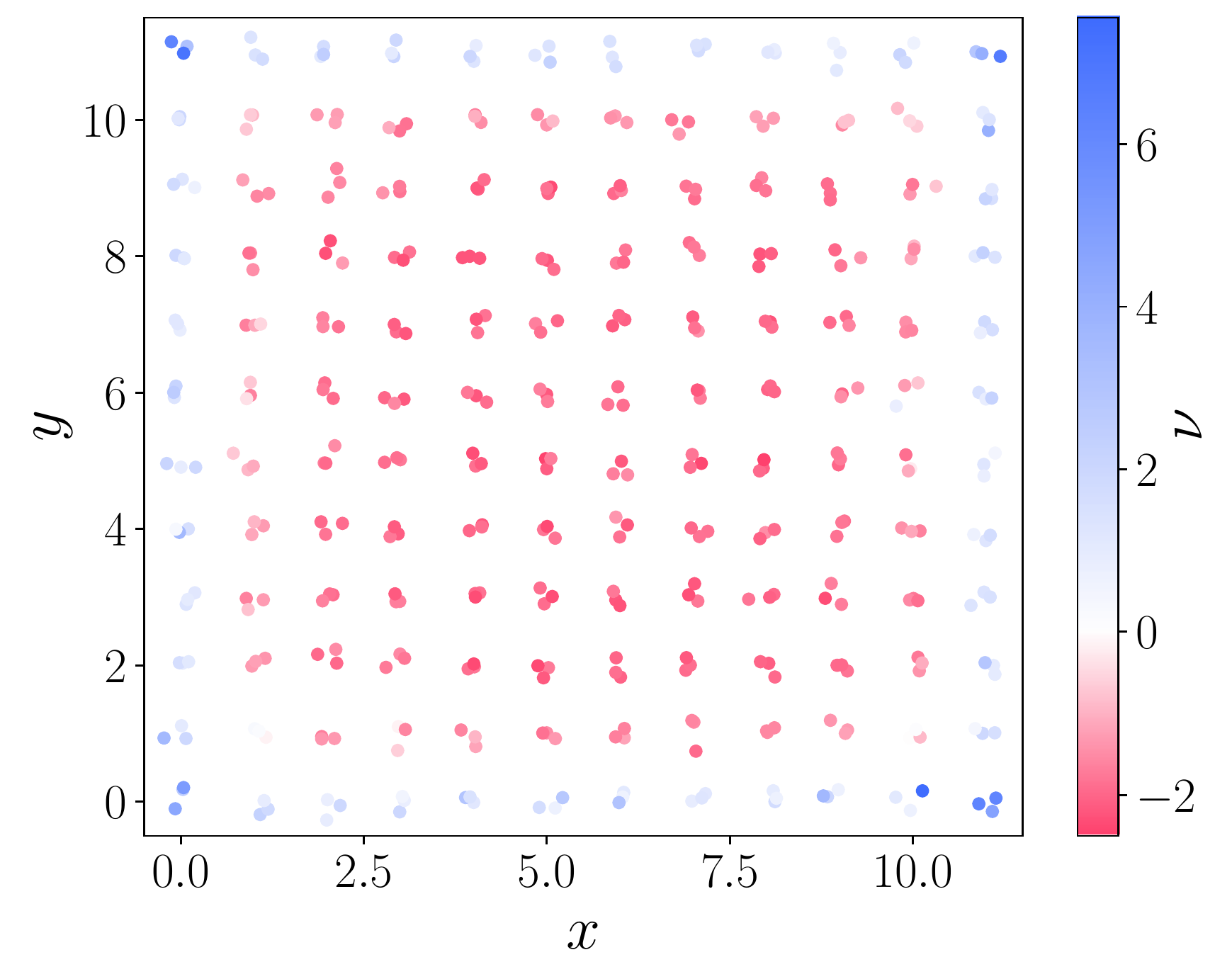}
    \caption{The chiral marker $\nu(\pos)$ for a single realization of the amorphous lattice with open boundary conditions. 
	The figure shows the projection into the $xy$-plane of the markers at all sites contained in a slice of width $1.5$ sites in the $z$ direction cut from the center of the full sample. 
	The system size is $L=12$, the width of the Gaussian distribution is $w=0.1$, and the parameters $M=0$, $t=0$ and $\lambda=1$ in the Hamiltonian Eq.~\eqref{app:eq:Hamiltonian}.}
   \label{app:fig:sup1}
\end{figure} 

\section{Averaging the local marker in systems with a boundary}

Averaging the Chern marker over the entirety of an open system leads to a vanishing result~\cite{bianco11}. 
This implies that the values of the Chern marker close to the boundary, where the state is non-local, must have a large value, scaling with the system size, so that they cancel the constant contribution from the bulk.
The same is true for the local markers presented in this paper.
The reason is that the summation of the marker over the entire open system is proportional to the trace of an operator, which for the chiral and Chern-Simons markers takes the form:

\begin{equation}
\label{app:eq:tr1}
\text{Tr}(\rho S\varepsilon^{i_{1},\dotsc,i_{n}}X_{i_{1}}\rho X_{i_{2}}\rho\dotsb\rho X_{i_{n}}).
\end{equation}
Implementing the anti-commutation relation $\{\rho,S\}=S$ to change the order of the first $\rho$ and $S$ in Eq.~\eqref{app:eq:tr1}, results in $\rho S$ being replaced by $S-S\rho$.
The first term yields
\begin{equation}
\text{Tr}(S\varepsilon^{i_{1},\dotsc,i_{n}}X_{i_{1}}\rho X_{i_{2}}\rho\dotsb\rho X_{i_{n}})=\text{Tr}(S\varepsilon^{i_{1},\dotsc,i_{n}}\rho X_{i_{2}}\rho\dotsb\rho X_{i_{n}}X_{i_{1}})=0,
\end{equation}
where the first equality follows from the cyclic property of the trace together with the commutation relation $[S,X_{i}]=0$, and the second follows
from the fact that the product $X_{i_{n}}X_{i_{1}}$ vanishes under anti-symmetrization. 
The remaining term is
\begin{equation}
-\text{Tr}(S\rho\varepsilon^{i_{1},\dotsc,i_{n}}X_{i_{1}}\rho X_{i_{2}}\rho\dotsb\rho X_{i_{n}})=-\text{Tr}(\rho\varepsilon^{i_{1},\dotsc,i_{n}}X_{i_{1}}\rho X_{i_{2}}\rho\dotsb\rho X_{i_{n}}S),
\end{equation}
 again using the cyclic property.
Commuting $S$ through all the $\rho$'s gives the equality
\begin{equation}
\label{app:eq:tr2}
\text{Tr}(\rho S\varepsilon^{i_{1},\dotsc,i_{n}}X_{i_{1}}\rho X_{i_{2}}\rho\dotsb\rho X_{i_{n}})=-\text{Tr}(\rho S\varepsilon^{i_{1},\dotsc,i_{n}}X_{i_{1}}\rho X_{i_{2}}\rho\dotsb\rho X_{i_{n}}),
\end{equation}
so that summing the marker over the entire open system indeed results in a vanishing result. 
The values at the boundary will therefore necessarily be large and grow with system size to cancel the contribution in the bulk.
Figure~{\ref{app:fig:sup1}} exemplifies this effect showing the projection into the $xy$-plane of the chiral marker at all sites contained in a slice of width $1.5$ sites in the $z$ direction, cut from the center of an amorphous superconductor in three dimensions with open boundary conditions.

Calculating the total marker numerically does not necessarily yield a vanishing result when $\rho$ is generated from the ground state of a Hamiltonian, even though this is the case theoretically.
Since the Hamiltonian for a topologically non-trivial phase is gapless on an open system, the ground state is not well defined.
Choosing a state randomly in the ground state manifold will in general not result in a state which obeys the constraint $\{\rho,S\}=S$ at the boundary, which therefore invalidates the argument above, as  Eqs.~(\ref{app:eq:tr1}-\ref{app:eq:tr2}) no longer hold.
In this case there is no reason for the total marker to vanish. 
This problem does not arise when calculating the Chern marker since there are no symmetry constraints to adhere to in this case.


%

\section{The choice of the projector $R$.}

In the main text we define the Chern-Simons marker in terms of the single particle density matrix for the chiral state:
\begin{equation}
Q =\frac{1}{2}\left(1+i|[\rho,S_{R}]|^{-1}[\rho,S_{R}]\right),
\label{app:eq:Qdef}
\end{equation}
where $|[\rho,S_{R}]|$ is the matrix absolute value, and $Q$ obeys a chiral constraint $S_{R}=1-2R$, such that $\{Q,S_{R}\}=S_{R}$.
As stated in the main text, $R$ can be chosen as an arbitrary trivial projector as long as $i[\rho,R]$ is gapped and the time-reversal constraint $TR^{*}T^{\dagger}=1-R$ is fulfilled.
In the amorphous system we use in the main text, several on-site projectors fulfill the time-reversal constraint where $T=i\sigma_{y}$, examples include $R=\frac{1}{2}(1-\tau_{y})$, $ R=\frac{1}{2}(1-\sigma_{x})$, $R=\frac{1}{2}(1-\sigma_{z})$; $R=\frac{1}{4}(1-\tau_{y})\otimes(1-\sigma_{y})$.
However, $R=\frac{1}{2}(1-\tau_{y})$ is the most natural choice both because it gives a gapped $i[\rho, R]$ in the parameter range we study, and it preserves the spin-rotation invariance that our model has after averaging over realizations. 

In the general case, there could be situations where no on-site operator fulfills the time-reversal constraint, or does not satisfy that $i[\rho,R]$ is gapped.
In that case, one has to consider more general trivial operators $R$.
The natural next step would be to consider an ansatz where $R$ is of the from 
\begin{equation}
R=\bigoplus_{A_{(\mathbf r,\mathbf r^\prime)}}R_{A_{(\mathbf r,\mathbf r^\prime)}},
\end{equation}
where $\oplus$ denotes direct sum, $R_{A_{(\mathbf r,\mathbf r^\prime)}}$ is an operator acting within the single-particle Hilbert space related to $A$ where $\{A\}$ are non-overlapping sets of nearest neighbors. 
If one cannot find a working $R$ of this form one has to move on and consider ans{\"a}tze $R$ where $A$ are not pairs of sites, but triplets, etc.

\end{document}